\documentclass[twocolumn,prl,showpacs]{revtex4}

\usepackage{graphicx}
\usepackage{dcolumn} % Align table columns on decimal point
\usepackage{subfigure}
\usepackage{times}

\def\simlt{\mathrel{\lower .3ex \rlap{$\sim$}\raise .5ex \hbox{$<$}}}
\def\simgt{\mathrel{\lower .3ex \rlap{$\sim$}\raise .5ex \hbox{$>$}}}

\def\NP{ {\rm NP} }
\def\P{ {\rm P} }

\def\and{ {\wedge} }

\begin{document}
\title{
Renormalization group approach to satisfiability}

\author{
S.N. Coppersmith}

\affiliation{Department of Physics, University of Wisconsin,
1150 University Avenue, Madison, WI  53706 USA}

\date{\today}

\begin{abstract}
Satisfiability is a classic problem in computational
complexity theory, in which one wishes to determine
whether an assignment of values to a collection
of Boolean variables exists in which all of
a collection of clauses composed of logical OR's of
these variables is true.
Here, a renormalization group
transformation is constructed and used to relate
the properties of satisfiability problems with different numbers
of variables in each clause.
The transformation yields new insight
into phase transitions delineating ``hard'' and
``easy'' satisfiability problems.

\end{abstract}

%05.45.-a Nonlinear dynamics and nonlinear dynamical systems
%05.10.Cc Renormalization group methods
%05.90.+m other topics in statistical physics, thermodynamics, and
%nonlinear dynamical systems
%89.20.Ff Computer science and technology
%02.60.Pn Numerical optimization
%75.10.Nr Spin-glass and other random models

\maketitle

Computational complexity theory addresses the question
of how fast the resources required to solve a given problem
grow with the size of the
input needed to specify the
problem.\cite{papadimitriou94}
$\P$ is the class of problems that
can be solved in polynomial time, which means
a time that grows as a polynomial of the size
of the problem specification,
while $\NP$ is the class of problems
for which a solution can be verified in polynomial time.
Whether or not P is distinct from NP has been a central
unanswered question in computational complexity
theory for decades.\cite{clay_website}

Satisfiability (SAT) is a classic problem in computational
complexity.
An often-studied type of SAT is  ${K}$-SAT,
in which one attempts to find assignment of ${N}$
variables such that the conjunction (AND) of ${M}$ constraints, or
clauses, each of which is the disjunction (OR) of ${K}$
literals, each literal being either a negated or un-negated variable, is true.
(This way of writing the problem, as a conjunction
of clauses that are disjunctions, is called conjunctive normal form.)
For example, the $3$-SAT instance with the
4 variables $x_1$, $x_2$, $x_3$, and $x_4$ and the four clauses
\begin{eqnarray}
& & \rm{(x_1=1 ~OR~ x_2=0 ~OR~ x_4=1)}\nonumber\\
\rm{~AND~ } & & \rm{(x_1=0 ~OR~ x_3=1 ~OR~ x_4=0)} \nonumber\\
\rm { ~AND~} && \rm{  (x_2=1 ~OR~ x_3=1 ~OR~ x_4=1)}\nonumber\\
\rm { ~AND~}  & & \rm{(x_1=0 ~OR~ x_3=0 ~OR~ x_4=1)}~,
\label{eq:simplest_sat_example}
\end{eqnarray}
is satisfiable because it it is true for
the assignments $x_1=1$, $x_2=1$, $x_3=1$, $x_4=1$.
Below, we will write satisfiability problems in conjunctive normal form
using the notation of \cite{satlib_website}, where the ANDs and ORs
are implied and the literals have positive or negative signs depending
on whether or not they are negated.  For example,
the expression of
Eq.~(\ref{eq:simplest_sat_example}) is written
\vskip .1cm

(1 ~ -2 ~ 4),~ (-1 ~ 3 ~ -4),~ (2 ~ 3 ~ 4),~ (-1~ -3 ~ 4)~.
\vskip .13cm
2-SAT can be solved in polynomial time~\cite{papadimitriou94},
while
${{K}}$-SAT with ${K} \ge 3$
is known to be NP-complete~\cite{cook71}:
if a polynomial
algorithm for $3$-SAT exists, then P is equal to NP.
The complexity of SAT is intimately related to
the presence of phase transitions%
~\cite{monasson99,kirkpatrick94,%
mezard02a,biroli00,mezard02b,mora05,achlioptas05}.
For random problems with $N$ variables, $M$ clauses,
and $K$ literals per clause, as
$M$ is increased there is a phase
transition from a satisfiable phase, in which almost
all random instances are satisfiable, to an unsatisfiable
phase, in which almost all random instances are
unsatisfiable.
The most difficult instances are near this
SAT-unSAT transition.
It has also been shown that there is a transition
as the parameter $K$ is changed
between $2$ and $3$, at $K_c\sim 2.4$,
at which the nature of the SAT-unSAT transition changes%
~\cite{monasson99}.

Here, we investigate the relationship between satisfiability
problems with different values of $K$ by constructing
a renormalization group transformation, similar
to those
used for phase transition problems~\cite{wilson70,wilson79,maris78,white92},
that reduces
the number of degrees of freedom, while possibly
increasing the number and range of interactions~\cite{niemeijer76}
(which in this context is the number of literals per clause).
To do this, we note that the expression
\begin{eqnarray}
&&((\mathcal{A}_1 \ x) ,~ (\mathcal{A}_2 \ x) ,~ \ldots (\mathcal{A}_P \  x),
\nonumber\\
&& (\mathcal{B}_1 \ -x) ,~ (\mathcal{B}_2 ~ -x))
\ldots ,
(\mathcal{B}_Q  \ -x)) \nonumber
\end{eqnarray}
is satisfiable if and only if
\begin{eqnarray}
&&((\mathcal{A}_1 \ \mathcal{B}_1) ,~ (\mathcal{A}_1 \ \mathcal{B}_2),\ldots
 (\mathcal{A}_1 \ \mathcal{B}_Q),\nonumber\\
&&~~(\mathcal{A}_2 \ \mathcal{B}_1) ,~ (\mathcal{A}_2 \ \mathcal{B}_2),\ldots
 (\mathcal{A}_2 \ \mathcal{B}_Q),\nonumber\\
&&~~\ldots,\nonumber\\
&& ~~(\mathcal{A}_P \ \mathcal{B}_1) ,~ (\mathcal{A}_P \ \mathcal{B}_2),\ 
\ldots ,~ (\mathcal{A_P} \  \mathcal{B_Q})) \nonumber
\end{eqnarray}
is satisfiable.
Here, the $\mathcal{A}_i$'s and $\mathcal{B}_i$'s are arbitrary clauses
and $x$ is a variable.
(The easiest way to see the equivalence is to note that both
expressions are satisfiable if and only
($\mathcal{A}_1$ AND $\mathcal{A}_2$ AND $\ldots$ $\mathcal{A}_P$) OR
($\mathcal{B}_1$ AND $\mathcal{B}_2$ AND $\ldots$ $\mathcal{B}_Q$) is.)
The first step of the renormalization procedure is to use this
identity to eliminate a given variable.
In this step, P clauses in which a given
variable comes in un-negated and Q clauses in which the same
variable comes in negated are eliminated and replaced
with PQ ``resolution''~\cite{robinson65} clauses.
Thus, eliminating a ``frustrated''~\cite{toulouse79}
variable (one that enters
into different clauses
negated and un-negated) increases the number
of clauses if PQ-(P+Q)$>$0.
The resolution of
two clauses of length $K_i$ and $K_j$ has
length $K_i+K_j-2$.
Note that resolving two 2-clauses yields a 2-clause,
resolving a 2-clause with a clause of length $\mathcal{K}\ge 3$
yields a clause length $\mathcal{K}$,
and resolving two clauses of with lengths $\mathcal{K}_1\ge 3$
and $\mathcal{K}_2\ge 3$ yields a clause with length greater than
both $\mathcal{K}_1$ and $\mathcal{K}_2$.

One then simplifies the resulting satisfiability expression by
noting that
\begin{enumerate} %(1)
\item
Duplicate clauses are redundant,
\item %(2)
Duplicate literals in a given clause are redundant,
\item % (3)
If a variable enters into one clause both negated and un-negated,
then the clause must be true and can be removed,
\item % (4)
If a clause has one literal, then the value of the corresponding variable
is determined, and
\item % (5)
If a subset of the literals in a clause comprise a different clause,
then the clause with more literals is redundant.
\end{enumerate}
This last point means, for example, that if an expression contains
both (1 3 -4 5) and (1 3), then (1 3 -4 5) can be removed,
because it is satisfied automatically
if (1 3) is satisfied.

This procedure is
known
in computer science
as ``the Davis-Putnam procedure of 1960~\cite{davis60} with
subsumption~\cite{cook79},"
and was originally proposed as
a method for solving satisfiability instances.
It does not perform well in practice~\cite{zhang02},
and has been proven to require exponential
time on some instances~\cite{tseitin68,haken85}.
However, here the aim is not to solve a given instance,
but rather to investigate the ``flow'' of the problem itself as
variables are eliminated~\cite{kadanoff66,wilson70}.
In particular,
this renormalization group (RG) transformation provides a natural framework
for understanding a phase transitions between ``easy'' and ``hard''
satisfiability problems identified in~\cite{monasson99}.

We present evidence that the change in the nature of the SAT-unSAT phase
transition critical value ${K_c}\sim 2.4$~\cite{monasson99}
is intimately related to whether or not the number
of clauses proliferates exponentially upon repeated
application of the renormalization group (RG) transformation.
Note that when $K=2$ the clause length decreases upon renormalization,
since the resolution of two 2-clauses
is a 2-clause, so no clause gets longer,
and some of the resulting clauses have a duplicate literal and so
get shorter.
Having a large number of 2-clauses limits the growth in the number
of long clauses because of subsumption,
so there is a qualitative difference in the behavior depending on
whether the ratio of the number of 2-clauses to the number of variables
grows or shrinks upon renormalization.

\begin{figure}[t]
\begin{center}
\vskip .4cm
%\subfigure{\includegraphics[height=1.8in]{1a.eps}}
%~\quad\subfigure{\includegraphics[height=1.8in]{1b.eps}}
\subfigure{\includegraphics[height=1.8in]{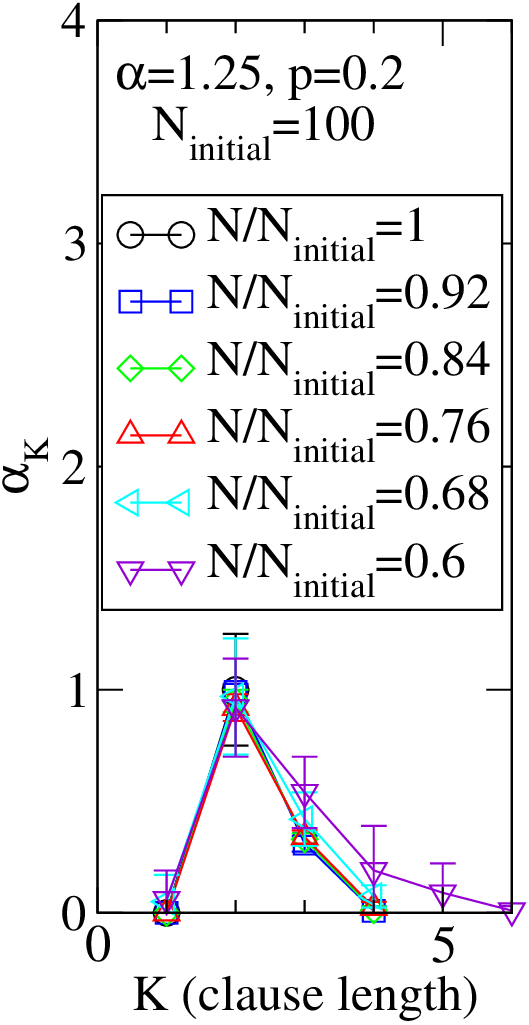}}
~\quad\subfigure{\includegraphics[height=1.8in]{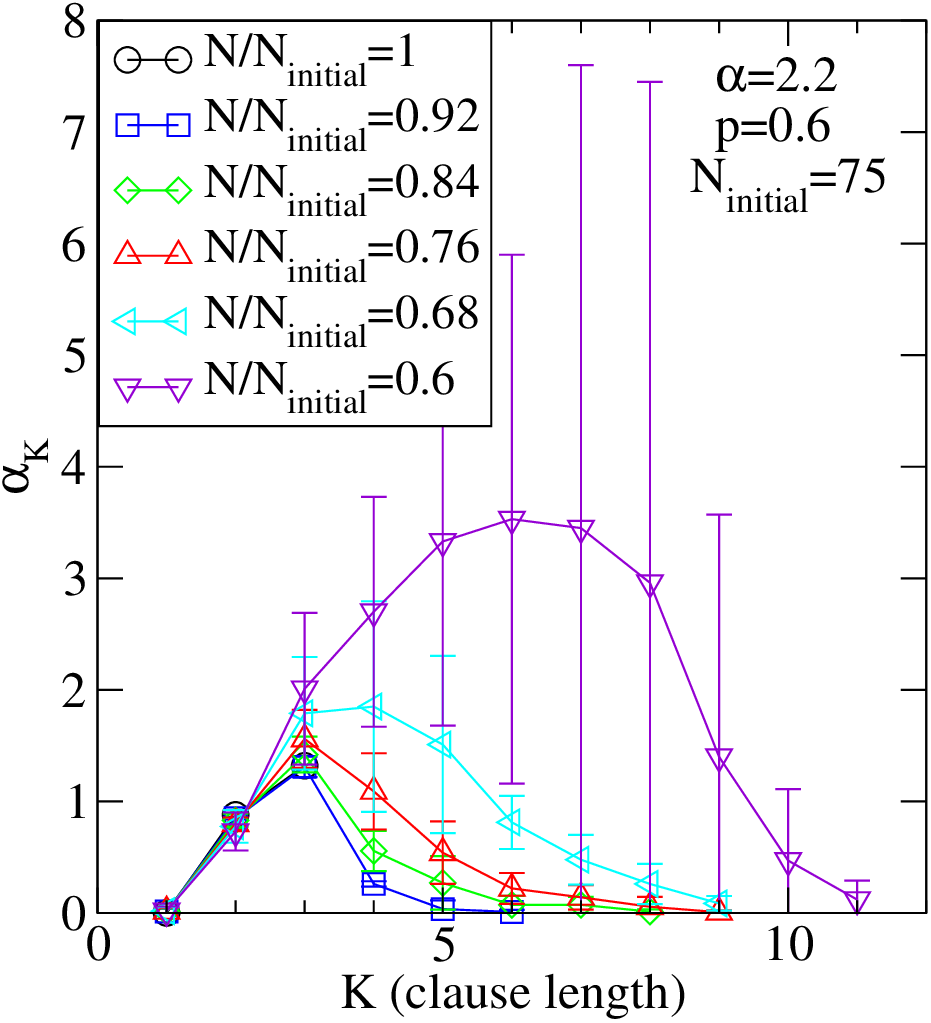}}
\caption{Plot of $\alpha_K$, the ratio of $M_K$, the number of clauses of
length $K$, to $N$, the number of variables, at different stages of the
renormalization process.
The points plotted are the mean and standard deviation of the results
from five independent system realizations.
The parameters are chosen to be at the SAT-unSAT transition
with $p=0.2<p_c$ (left panel) and $p=0.6>p_c$ (right panel).
When $p>p_c$ the clause length increases markedly and the number of
clauses grows enormously.
}
\label{fig:t_dependence}
\end{center}
\end{figure}

We show numerical data for an RG implementation in which
successive variables are chosen randomly and eliminated if they occur
in a clause of minimum length.
This procedure is used because it focuses on
short clauses, which are much more restrictive
than long clauses.
Figure~\ref{fig:t_dependence} shows
$\alpha_K$, the ratio of $M_K$,
the number of clauses of length $K$ to $N$, the number of variables
remaining in the problem,
as a function of $K$, as the RG proceeds. 
The average and standard deviation of numerical
data from 5 realizations at the SAT-unSAT
transition with $p=0.2$ and $p=0.6$ are shown
(using parameter values for the transition locations
from~\cite{monasson99}).
Large numbers of long clauses are generated when
$p=0.6>p_c$ and not when $p=0.2<p_c$.
%\begin{figure}[htbp]

\begin{figure}[t]
\begin{center}
\vskip .4cm
%\subfigure{\includegraphics[height=1.75in]{2a.eps}}
%~\quad\subfigure{\includegraphics[height=1.75in]{2b.eps}}
\subfigure{\includegraphics[height=1.75in]{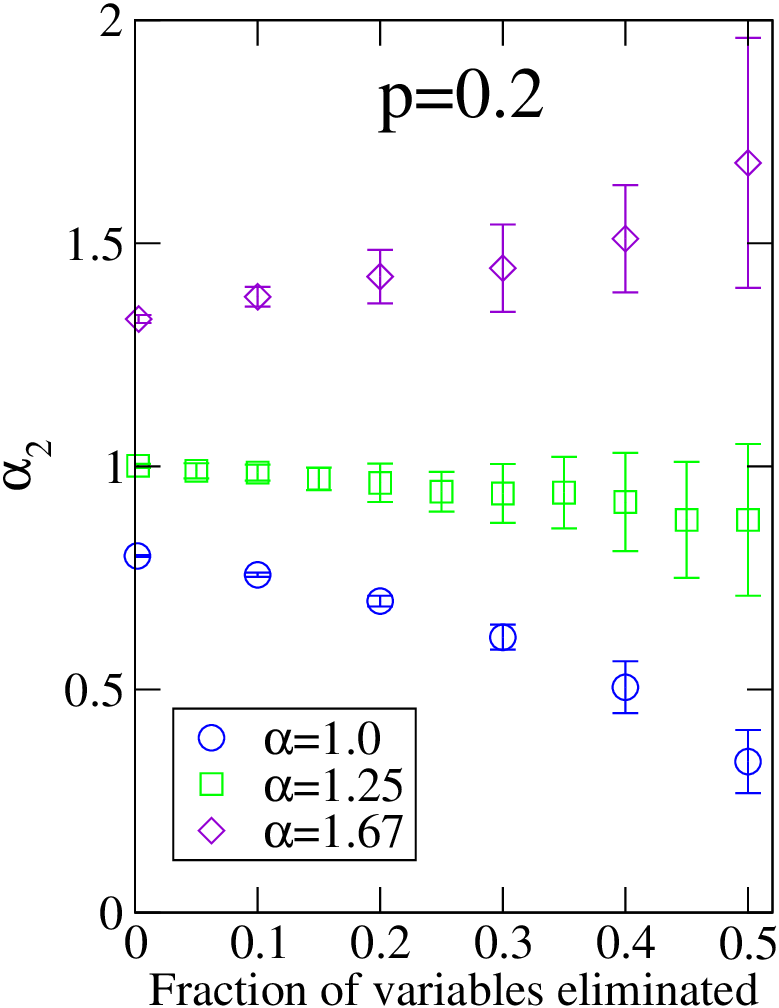}}
~\quad\subfigure{\includegraphics[height=1.75in]{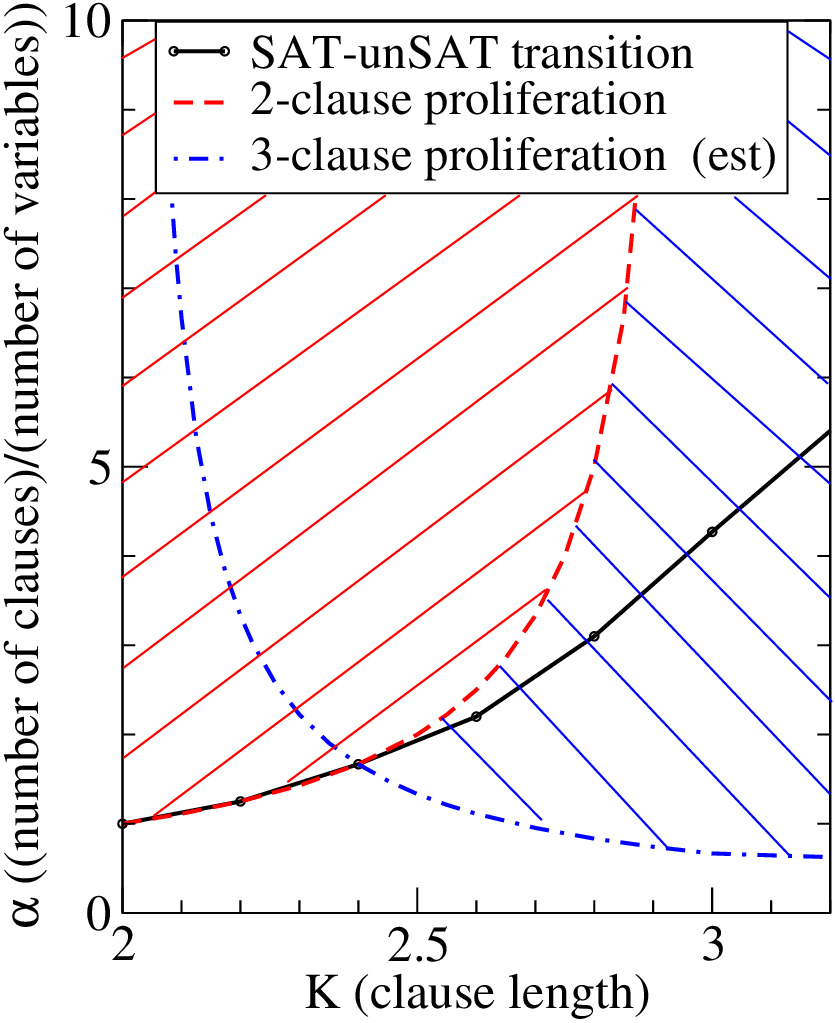}}
\caption{Left: plot of $\alpha_2$, the ratio of the number of 2-clauses
to the number of variables, versus $N$, the number of undecimated
variables, for instances with $p=0.2$ and different initial values of $\alpha$.
The numerical data are averages and standard deviations of five
realizations of systems of size 500 when the initial $\alpha=1$,
size 400 when the initial $\alpha=\alpha_c=1.2$, and size
300 when the initial $\alpha=1.67$.
The numerical data are consistent with the hypothesis that when
$K<K_c$ the SAT-unSAT transition occurs when $\alpha_2$ neither
decreases nor increases upon renormalization.
Right: Schematic phase diagram showing the SAT-unSAT transition
(using data of Ref.~\protect{\cite{monasson99}}), the region in
which the number of 2-clauses increases upon renormalization (the red
hatched region in the left of the figure)
and an estimate of the region in which the number of clauses with $K\ge 3$
increases upon renormalization (the blue hatched region in the right
of the figure).
The SAT-unSAT transition line crosses into the region in which long
clauses proliferate exponentially at the intersection of the three
lines.  The estimate for 3-clause proliferation given in the
text yields an intersection at ($K=2.4$, $\alpha=1.25)$.
}
\label{fig:figure2}
\end{center}
\end{figure}

It has been proven that the SAT-unSAT transition for 2-SAT occurs
at $\alpha=1$~\cite{bollobas01}, and
for $2\le K<2.4$, the SAT-unSAT transition is believed to occur
when $\alpha_2=1$~\cite{monasson99}.
Figure~\ref{fig:figure2} (left) shows that when $K=2.2$, $\alpha_2$ increases
upon renormalization when $\alpha_2>1$ and decreases
upon renormalization when $\alpha_2<1$.
Because adding additional
three-clauses does not affect the behavior of
the two-clauses,
and because two-clauses are much more restrictive than longer clauses,
the two-clauses dominate the problem whenever $\alpha_2>1$.
One would expect that 3-clauses by themselves would
proliferate when $N_{initial}$, the initial number of variables, was equal to 
$3 M_3 /2$.  (This estimate, the analog of the result for two-clauses, follows
from setting the number of literals in all 3-clauses to twice the number of variables,
which means that on average each variable enters into one 3-clause negated
and one 3-clause un-negated.
Moreover, eliminating one variable on average yields two less literals and one
less variable, so that, ignoring fluctuations, the relationship remains true.)
However, because the 2-clauses prevent the 3-clauses from proliferating,
adding 3-clauses to the 2-clauses changes the nature of the
SAT-unSAT transition only when enough 3-clauses have been added
so that the SAT-unSAT transition occurs with
$\alpha_2<1$.
In this regime, under renormalization the 2-clauses disappear and
so the clauses all become longer.
Since very long clauses are ORs of
many literals and hence easy to satisfy, the number of clauses
must go up sufficiently fast for the problem to be difficult
to solve---at
large $K$, the SAT-unSAT transition
occurs when the ratio of the number of clauses to the
number of variables is $\propto 2^K$~\cite{achlioptas05}.
When $K>K_c$, near the SAT-unSAT transition
we expect the number of clauses to grow geometrically with
iteration number, and
the numerical data are consistent with the maximum number of
clauses obtained during the renormalization process
increasing exponentially with  the initial number of variables,
with an exponent that increases monotonically with $(M/N)_{initial}$,
the initial ratio of the number of clauses to the number of variables.
Figure 2 (right) shows a phase boundary lines for the SAT-unSAT transition,
for the onset of increase in the number of 2-clauses, and our estimate
for the onset of
proliferation of clauses of length greater than or equal to three.
%\begin{figure}[htbp]

When $K>K_c$, the SAT-unSAT transition%
~\cite{monasson99,kirkpatrick94,%
mezard02a,biroli00,mezard02b,mertens03,mora05}
occurs in a regime in which
the renormalization transformation causes both
the typical clause length and the total number of
clauses to grow.
We conjecture that the SAT-UNSAT transition
occurs at the value of $M/N$ at which
the rate of exponential growth is a critical value.
A ``replica-symmetry breaking"
transition~\cite{mezard02a,mezard02b,mertens03,mora05}
at a somewhat smaller value of $\alpha$ can be interpreted
in terms of propagation of constraints on eliminated literals,
as will be discussed elsewhere.~\cite{coppersmith05b}
Because of the exponential clause proliferation,
numerical investigation of these transitions
using this renormalization group is limited to small sizes.
However, the renormalization group may still be useful for
investigating these transitions using
analytic techniques appropriate
for large $K$~\cite{mora05,achlioptas05} for any $K > K_c$,
though it will be necessary to understand how to account for
possible RG-induced
correlations between clauses.

When $K>K_c$, the number of clauses continues to increase
under renormalization
until it is no longer unlikely that
a given compound clause contains a repeated variable
(in addition to the decimated one),
which we expect to occur when
the renormalized clause length
is of order $\sqrt{{N}}$~\cite{diaconis89}.
Because it appears that there is no impediment to the
growth in the effective value of $K$ until it is of
order $\sqrt{N}$, where
the problem specification itself is exponentially
large in $N$, it appears that the renormalization
group procedure can
transform the problem
out of NP and even PSPACE altogether.
This property may indicate that whether or not a given
computational problem
has a solution that can be verified using polynomially-bounded
resources has no fundamental effect on the difficulty of solving
the problem.

In summary, a transformation inspired
by the renormalization group is constructed and used
to relate the behavior of
satisfiability problems
with different values of ${K}$, the number of literals per clause.
The transformation provides useful insight into previously
identified phase transitions of satisfiability problems
and may yield new insight into the question of whether
or not $\P$ is equal to $\NP$.

%Acknowledgments
The author
gratefully acknowledges financial support from grants
NSF-DMR 0209630 and NSF-EMT 0523680, and
the hospitality of the Aspen Center for Physics,
where some of this work was done.

%\bibliography{/Users/suecoppersmith/Documents/papers/bibfiles/prlbib}
\bibliography{prlrevisedbib}

\end{document}